# LOW-ENERGY RUN OF FERMILAB ELECTRON COOLER'S BEAM GENERATION SYSTEM*

L. R. Prost#, A. Shemyakin, FNAL, Batavia, IL 60510, U.S.A.
A. Fedotov, J. Kewisch, BNL, Upton, NY 11973, U.S.A.

## Abstract

In the context of the evaluation of possibly using the Fermilab Electron Cooler for the proposed low-energy RHIC run at BNL, operating the cooler at 1.6 MeV electron beam energy was tested in a short beam line configuration. The main conclusion of this feasibility study is that the cooler's beam generation system is suitable for BNL needs. The beam recirculation was stable for all tested parameters. In particular, a beam current of 0.38 A was achieved with the cathode magnetic field up to the maximum value presently available of 250 G. The energy ripple was measured to be 40 eV. A striking difference with running the 4.3 MeV beam (nominal for operation at FNAL) is that no unprovoked beam recirculation interruptions were observed.

## INTRODUCTION

Electron cooling proposed to increase the luminosity of the RHIC collider for heavy ion beam energies below 10 GeV/nucleon [1] needs a good quality, 0.9-5 MeV electron beam. Preliminary design studies indicate that the scheme of the Recycler's electron cooler at FNAL is suitable for low-energy RHIC cooling and most parts of the cooler can be re-used after the end of the Tevatron Run II. To analyze issues related to the generation of the electron beam in the energy recovery mode and to gain experience with the beam transport at lower beam energy, a dedicated study was performed at FNAL with a beam run through a short beam line (so called U-bend). This report summarizes our findings and observations in the course of the measurements.

## SETUP

The Pelletron [2] is a 6MV electrostatic accelerator which works in the energy recovery mode. It consists of two acceleration columns contained in a pressure vessel filled with pressurized SF6 gas (~70 psi) (Fig. 1). Focusing and steering is provided by solenoids each with a pair of dipole correctors. The electron gun is embedded in a magnetic field at the terminal and can generate a few Amperes electron beam. A collector recuperates the beam which is decelerated to 3 kV. In the U-bend configuration the beam generated in the gun goes straight down to the 180° bend (U-Bend) and returns to the collector. One important feature for tuning purposes is the lack of Beam Position Monitors (BPMs) inside the acceleration and deceleration columns. The first BPM is located right at the exit of the acceleration column; the last BPM just before the entrance of the deceleration tube. More details on the Pelletron and its operation can be found elsewhere (for instance in Ref. 3).

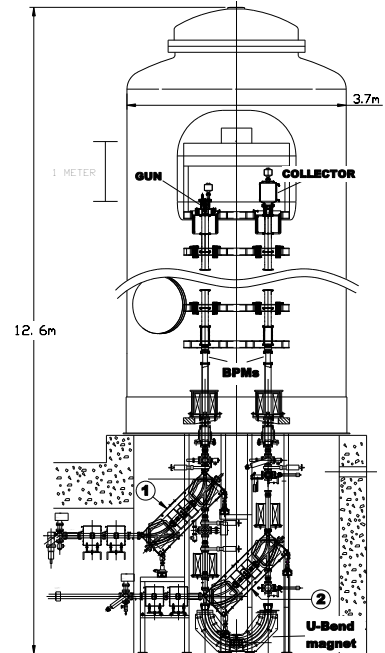

Fig. 1 Sketch of the Pelletron in the U-bend configuration. 1&2 point to the 90° bending magnets (each composed of two 45° bending magnets + a solenoid in between) that are used to circulate the beam through the cooling section and back to the collector.

## HIGH VOLTAGE PERFORMANCE

For the measurements presented in this paper, the Pelletron was operated at 1.6 MV. Because it is conditioned to 5 MV for normal operation, high voltage (HV) discharges without beam were not a concern. Thus, the Pelletron performance is solely characterized by the HV ripples.

### High voltage regulation system

The HV regulation system is described in detail in Ref [4]. In short, the terminal voltage is measured by a Generation Voltmeter (GVM) and its value is compared with the set point. An error signal is generated and brought to the terminal where a corresponding current is emitted from a set of needles protruding from the terminal shell toward the tank wall. Performance of the HV regulation circuitry can be characterized by increasing the chain current with other parameters being fixed (Fig. 2). For a low chain current, the HV is below the set point, and the regulation circuitry suppresses the needles current. As the result the HV increases linearly with the chain current. The deviation from linearity between the HV and the chain current at the beginning of the curve is due to the terminal-to-ground capacitance when increasing the chain current too rapidly. When the HV is close to the set value, the regulation circuitry adjusts the needles current to be roughly equal to the increase of the chain current. At 1.6 MV, the slope of the I-V curve in the region of regulation gives an effective resistance of ~90 MOhm as for 4.3 MV [4]. Therefore, one should expect to have the same suppression of the chain current fluctuations as well.

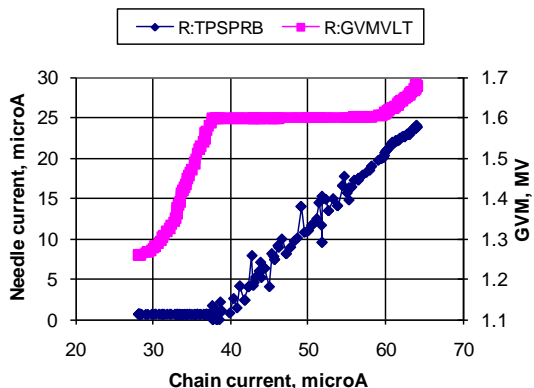

Fig. 2 Terminal voltage and needles current as a function of the chain current.

When the chain current increased to ~60 μA, the needles current reaches its maximum in the regulation mode, and the HV starts to increase faster again. Consequently, the maximum needles current with good regulation is 15μA. Most of the measurements were made at the significantly lower current of 2 μA (on average).

This maximum needles current at a given HV is determined by the electronics design and the needles position with respect to the terminal shell. For operation at 1.6 MV, the needles were moved to the most outward position, 25 mm, resulting in a maximum possible needles current of ~15μA. An attempt to run the Pelletron at even lower HV, 1 MV, showed that at 25 mm the maximum needles current in the regulation mode drops to ~1 μA, which does not allow operation with beam using the present configuration. If operation at 1 MV is needed, modifications to either the needle motion system or the terminal electronics will be necessary.

*Energy ripple*

The chain current, which fluctuation is the main source of the energy ripple, went down to roughly the same proportion as the HV with respect to nominal operation at 4.3 MV. Therefore, one should expect to have the relative HV ripple to be independent of the HV, and the value of ~100 V for the ripple found in [4] gives a ~40 V ripple (sigma) at 1.6 MV. A beam-based estimation of the energy ripple can be done from spectra of BPM signals (Fig. 3).

Comparison of spectra in the high- and low-dispersion part of the beam line shows that at frequencies < 5 Hz, the beam motion is caused primarily by HV fluctuations. By measuring the dispersion coefficients at the first BPM following the U-bend, applying a 0.5 -6 Hz filter and assuming that the entire signal comes from energy ripples, the rms ripple is estimated to be 38 eV, in good agreement with the constant relative HV ripple argument.

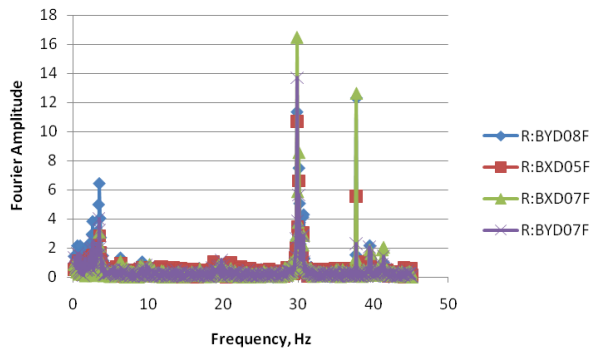

Fig. 3 FFT spectra of 4 BPMs in high-dispersion locations. The acquisition frequency was 81 Hz. For FFT in each channel, 1024 recorded points were used. The beam current was 100 mA.

## RECIRCULATION STABILITY

Prior to the study, optics simulations were carried out using OptiM [5]. From this exercise, we verified that known aperture restrictions (in the BPMs and accelerating columns) were not a concern to pass the beam to the collector at 1.6 MV. While the direct implementation of the settings from simulations did not allow circulating the beam without losses, they were a good starting point for further tuning. Note that simulations using the final settings arrived at during the measurements did not show any contradictions with experimental results *i.e.* envelopes larger than or close to the vacuum chamber along the beam line while the beam was transported cleanly to the collector in the experiment.

Stability of the beam recirculation was excellent in comparison with operation at 4.3 MV. There was not a single full discharge or unprovoked beam interruption. All interruptions were characterized by the terminal voltage slowly decreasing (*i.e.* becoming more positive) because of losses induced by tuning; the cathode current stayed almost constant until the protection system was detecting the decrease of HV and turning the beam off.

When a 0.1 A beam was left running it stayed uninterrupted for 20 hours and was stopped intentionally.

## MAXIMUM BEAM CURRENT

The values of the maximum DC current achieved in various configurations during the 1.6 MV run are summarized in Table 1.

In Case A, the maximum current was limited by a sharp growth of the beam losses that did not depend on focusing when the control electrode, which determines the total amount of current extracted, approached 0 kV. It was interpreted as an onset of the emission from the side (cylindrical) surface of the cathode. According to gun simulations, trajectories of electrons emitted from the side surface differ dramatically from the main beam, and these electrons are lost. This effect determines the maximum gun current at a given anode voltage.

Table 1: Maximum beam current achieved for different gun conditions

| Case # | Cathode field, G | Anode voltage, kV | Max. DC current, A |
|---|---|---|---|
| A | 84 | 10 | 0.24 |
| B | 253 | 10 | 0.20 |
| C | 255 | 20 | 0.38 |

In Case B, the process of increasing the beam current was stopped at 0.2 A, which was the initial target value. Because for both A and B, the maximum achievable current is 0.24 A, to prove that there is no immediate "hard" limit to the maximum current that can be recirculated at 1.6 MV, the anode voltage was increased (Case C), hence the maximum current that can be theoretically achieved. Reaching a beam current of 0.38 A was deemed a proof of this statement.

## TRANSVERSE ANGLES

The results presented in Fig. 3 showed that a large component of the beam motion is not associated with the energy ripple. The strongest lines in Fig. 3 are the rotation frequency of the shaft motor, 29.8 Hz, and its second harmonics, and the second harmonics of the chain motor rotation frequency, $2 \times 19.3 = 38.6$ Hz. The contribution can come either from vibrations (discussed in detail in Ref. [6]) or can be caused by stray magnetic fields from the motors.

An accurate estimation of a possible effect of these oscillations on angles in the cooling section requires tracking specific lines from the spectrum similarly to what was done in Ref. 6. To estimate the order of magnitude, we can assume that the beam sizes at the exit of the acceleration column location and in the proposed BNL cooling section are similar, and, correspondingly, the oscillation amplitudes will be similar as well. The resulting angles should be comfortably below the expected total angle of 0.1 mrad.

## HYSTERESIS IN BENDING MAGNETS

One of the concerns of using Fermilab's bending magnets for low-energy running is the quality of their magnetic fields at low field strength. For the U-bend mode of operation, the two 90° bending magnets (1&2 on Fig. 1) are turned off. Because two of the four 45° magnets that compose the 90° bends are located in the U-bend beam line, their hysteresis property can be estimated. For this purpose, the beam trajectory was recorded before and after cycling the bends from zero to their full nominal current (for 4.3 MeV operation) of ~4A and back to zero. Using the OptiM program, the field of the dipoles was fitted to match the resulting orbit differences measured in the BPMs after this cycle. From this, the fitted field integrals are calculated (Table 2).

Table 2: Change of the integrated dipole field strength calculated from the change in the orbit.

| | Acceleration side bend | Deceleration side bend |
|---|---|---|
| ∫$B_x$ [G m] | -0.305 | -0.555 |
| ∫$B_y$ [G m] | 0.009 | 0.067 |

When the magnets are used at the energy of 1.6 MeV, the observed field perturbation is ~0.5% of the main field. If the field quality is of the same order, it may result in focusing aberrations.

Measuring the magnet properties at low bending fields is desirable as well as foreseeing bipolar bend power supplies for degaussing. Note that the effect becomes even a bigger concern for running at lower energies.

## CONCLUSION

The low-energy run of Fermilab's Electron cooler showed that the system of beam generation and energy recovery is capable of operating at 1.6 MeV and should be able to deliver an electron beam with the appropriate properties for cooling. The beam transport at the required current did not present significant problems, and the recirculation stability was excellent. Nevertheless, several issues were identified:
- High voltage regulation does not work properly at $\lesssim 1$ MV; modifications (likely, minor) would be required.
- At the lower energy, the present protection system based on ionization chambers is inadequate.
- Additional magnetic measurements of the bending magnets are needed to determine at what parameters they can be used in the low-energy mode.

In summary, using the Electron cooler for the BNL low-energy RHIC program is feasible.